# REGIONAL PILOT STUDY TO EVALUATE e-READINESS AND LOCAL e-GOVERNMENT SERVICES

Marjan Angeleski[1], Pece Mitrevski[2], Slavica Rocheska[3] and Ane Lashkoska[4]

[1]University of St. Clement Ohridski, Faculty of Economics, Prilep, Macedonia

[2]University of St. Clement Ohridski, Faculty of Technical Sciences, Bitola, Macedonia

[3]University of St. Clement Ohridski, Faculty of Economics, Prilep, Macedonia

[4]Committee on Equal Opportunities for Women and Men, Assembly of the Republic of Macedonia

*ABSTRACT*

*The concept of local e-Government has become a key factor for delivering services in an efficient, cost effective, transparent and convenient way, in circumstances where a) citizens do not have enough time available to communicate with local authorities in order to perform their responsibilities and needs, and b) information and communication technologies significantly facilitate administrative procedures and citizens-government interaction. This paper aims to identify e-services that local authorities provide, and to investigate their readiness for delivering these services. A pilot research has been conducted to identify the offer of e-services by local authorities, along with e-readiness in municipalities of the Pelagonia region in the Republic of Macedonia. The survey was carried out by means of structured interview questions based on a modified model proposed by Partnership on Measuring ICT for Development – web analysis of municipal websites in the region has been conducted, as well. The study reveals uneven distribution according to the age group of users, lack of reliability and confidence for processing the needs and requests electronically by a large part of the population, and improperly developed set of ICT tools by local governments for providing a variety of services that can be fully processed electronically.*

## 1. INTRODUCTION

The ICT advancements in the last two decades have significantly affected all the domains of social life and, in this context, the functioning of the central and local governments, as well. The ICT development enabled governments to implement new ways for delivering services to citizens, businesses and government agencies. Similarly to the concept of electronic commerce, which allows better and more efficient communication between business partners electronically (B2B) and between businesses and their consumers (B2C), the concept of e-government enables more convenient, transparent and inexpensive interaction between government and citizens (G2C), government and businesses (G2B), and internally between different government departments (G2G) [1]. It is important to make a clear distinction between the terms e-government and e-governance. The concept of e-governance is much broader concept which "allows citizens to communicate with government, participate in the governments' policy-making and citizens to communicate each other and to participate in the democratic political process" [2]. Recent research indicates that it is about reinventing the way in which government interacts with citizens, governmental agencies, businesses, employees, and other stakeholders. It is about enhancing democratic processes and also about using new ideas to make lives easier for the individual citizen by, for example, transforming government processes, enabling economic development, and renewing the role of government, itself, in society [3]. On the other hand, the e-government can be defined as "utilizing the Internet and the world-wide-web for delivering government information and services to citizens" [4]. According to OECD, e-government includes "use of new information and communication technologies (ICTs) by governments as applied to the full range of government functions. In particular, the networking potential offered by the Internet and related technologies has the potential to transform the structures and operation of government" [5]. Hence, e-government means "using the tools and systems made possible by Information and Communication Technologies (ICTs) to provide better public services to citizens and businesses" [6].

Apparently ICT provides many opportunities for government, such as: (1) increased operational efficiency by reducing administrative costs and increasing productivity, (2) better quality of services provided by government agencies, [7], (3) increase responsibility, (4) more accurate delivery of services (5) reduction of time especially for repetitive administration tasks, etc.

On the other hand, E-government readiness assessment (ERA) evaluates how ready a country, a city, or even local government organization is to develop e-government, in order to assess its stage of readiness, identify its gaps, and then redesign its e-government strategy. Particularly, e-readiness assessment can help developing countries to measure and plan for ICT integration – it can help them congregate their efforts internally, and identify areas where external support is required [8]. E-Government readiness is a function of a country's state of net-worked readiness, its technological and telecommunication infrastructure, the level of citizen's access to electronic services and the existence of governmental policy and security mechanisms. [9]

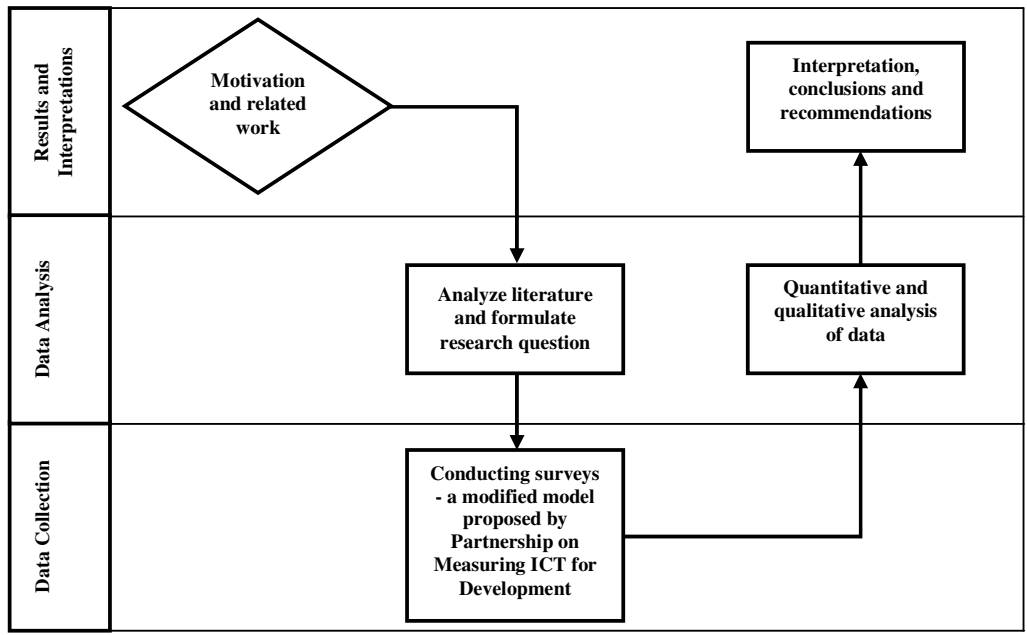

Figure 1. Research process flowchart

The remainder of this paper is organized as follows. Section 2 focuses on related work and identifies the main motivations. Section 3 discusses methods and instruments used in the pilot study, i.e. the methodology to investigate supply of e-services provided by the central government, which was modified and adjusted to the level of supplying services by local authorities (Fig.1). The principal findings are discussed in Section 4, where we shift our attention to the comparison between rural and urban municipalities, and the analysis of municipality e-services as a key indicator for e-government readiness. Finally, Section 5 concludes the paper.

## 2. MOTIVATION AND RELATED WORK

According to Hiller and Belanger [10] the implementation of e-government concept comprises five stages based on the levels of technology and sophistication: (1) Information Stage – at this stage, an informational site would simply post information for its constituents. (2) Two-way Communication stage – would allow the exchange of information through e-mails and feedback forms. (3) Transaction Stage website allows users to pay taxes online, renew licenses, etc. (4) Integration Stage – at this stage, all government services are integrated and which can be accomplished with a single portal so that its constituents can access the services no matter which agencies or departments offer them. And the last one, (5) Participation Stage – calls for the highest sophistication and is implemented by only a rare few. At this stage, the user could vote online or file comments online. These sites require a very high level of security and privacy and are very much in its infancy throughout the world.

The e-government contributes for improved efficiency, greater involvement of citizens, better quality of services and increased transparency. Even though the e-government incorporates a potential to provide many benefits [11], it is evident that there is insufficient research related to the analysis of e-government performance at the local level [12]. The term local e-government refers to the ICT usage by local authorities for offering services to citizens, i.e. one-stop access to multilevel local government services. According to Lanvin and Lewin [13] apart from typical national e-government services, local e-governments also attend to a large number of citizens' needs. Specific e-government services are increasingly handled at the local rather than national level. Hence, recent studies are increasingly focused on the analysis of e-government performance at the local level. They note that modern communication technologies can be especially effective in local e-government, creating added value for administration and users as well [14]. The Internet provides a powerful tool for reinventing local governments. It encourages transformation from the traditional bureaucratic paradigm, which emphasizes standardization, departmentalization, and operational cost-efficiency, to the "e-government" paradigm, which emphasizes coordinated network building, external collaboration, and customer service [15]. This transformation implies greater openness and responsiveness of local e-government to the needs and demands of citizens and businesses. Consequently, it allows easier access and effectiveness in the provision of local services and enhanced integration between local authorities and members of the local community.

In the last decade Macedonia made significant efforts to follow the general trends and to ensure broad implementation of the e-government concept. This strategic orientation resulted in development of infrastructure, management and administrative capacity for delivering services and ensuring effective communication and interaction with citizens and businesses. More intensive activities in the implementation of the concept of e-government were carried out after 2005 when a strategy for e-government and more regulations were adopted. A project about local e-government has started in 2006 covering 50 municipalities and has intended to provide wireless internet access, better services and greater ICT usage by citizens. The project has contributed for modernization of IT infrastructure and better internet connectivity. Consequently, a significant growth in electronic services supply by local government has been recorded. Certainly these changes, to a large extent, are the result of the improvement of infrastructural, technological and administrative conditions for implementation of e-government at local level and increased number of internet users. Thus, according to The Global Information Technology Report [16], that measures the Government Online Service Index, i.e. assesses the quality of government's delivery of online services, Macedonia is ranked at the 82 place on a 0-to-1 scale, out of 140 countries, with index 0.45.

In a similar effort, Todevski et al. [17] performed an analysis of the financial aspects of introducing one stop shop services in the Republic of Macedonia by using computer based information systems. The analysis was conducted using public data for the administrative services which are currently provided by a closed set of 16 Macedonian government institutions. They calculated the financial implications on citizens, businesses, institutions, and other entities in the society, i.e. the overall savings for the society, which can be used by decision-makers in order to adjust the degree of investments in information systems and other complementary assets needed for introduction of these services.

## 3. METHODS AND INSTRUMENTS

In this case, a pilot research has been conducted in 9 municipalities of the Pelagonia Region in the Republic of Macedonia in order to identify the offer of e-services by local authorities and their e-readiness. According to the Nomenclature of territorial units (third level adopted by the State Statistical Office in compliance with the classification of the European Union - Nomenclature of Territorial Units for Statistics – NUTS), Republic of Macedonia is divided into eight statistical regions: Pelagonia, Vardar, Southeast, Southwest, Skopje, Northeast, Polog, and East. Pelagonia region is located in the southern part of the country bordering with Greece and Albania. This region embraces 232,959 inhabitants (estimations for 2012, Regions of the Republic of Macedonia, 2013, State Statistical Office of the Republic of Macedonia) [18] representing 11.3 percent of the total population and is the largest region comprising 19.9 percent of the territory of the Republic of Macedonia. Pelagonia region covers nine municipalities (10.7 percent of the total number of municipalities) out of which 5 are urban municipalities (Bitola, Prilep, Resen, Krusevo and Demir Hisar) and 4 are rural municipalities (Krivogashtani, Dolneni, Mogila, Novaci) (Law on territorial organization of local government in the Republic of Macedonia, "Official Gazette of the Republic of Macedonia" no. 55/2004, 12 /2005, 98/2008 and 106/2008) [19]. In fact, it includes 343 inhabited places representing 19.4 percent of overall inhabited places in the Republic of Macedonia. In this region, in 2012, 54 percent of population aged 15 to 74 years used a computer and 53 percent used the Internet. The percentage of households that used a computer and Internet was 61 percent and 55 percent, respectively [18].

The research was conducted in two phases aimed at collecting primary and secondary data. In the first phase a questionnaire consisting of 7 closed-ended questions was used for collecting primary data. Further, the second phase encompassed a web research of the municipalities' official web pages in order to scan and analyze e-services offered by municipalities as a key indicator for e-government readiness. For each indicator of e-government, a readiness grade of sophistication is given through descriptive interval scale from 0 to 4, where "grade 0" means that the municipality does not offer specific electronic service and "grade 4" means that the particular service is completely web based. This categorization is based on the model included in Capgemini report [20] that measures the online availability of public services in Europe:
(Level 0) The service is not available electronically;

(Level 1) There is information available on the web site about the way in which citizens can obtain services by going directly to the municipality;

(Level 2) The municipality web site contains forms/templates for downloading that can be completed by citizens as a prerequisite for obtaining a specific service. This method reduces the time needed because the forms can be filled out at home;

(Level 3) An interactive service for getting certain services is implemented on municipality web site. This means that citizens can fill out the predefined web forms and submit them online. If a payment of certain fee for the required service is needed, it will not be included into the offered service but it will be realized additionally;

(Level 4) Complete internet services - the municipality web site includes interactive service for obtaining certain services where through predefined web forms citizens can fill out the necessary data online and submit them online. If payment of a service fee is required, it will be simultaneously made through the installed municipality system or by using other means e.g. SMS. Each citizen that has submitted a request will receive a response via the internet (e.g. via e-mail or tracking system of the subject), by phone or SMS. The decision regarding the request after notification may be personally taken at municipality rooms or may be sent by mail.

The methodology proposed in the Framework for a set of e-government core indicators is used for designing a questionnaire. Since this methodology is to investigate the supply of e-services provided by the central government, it was modified and adjusted to the level of supplying services by local authorities. Designing questions about the supply of services by local authorities is based on the previous research and systematization of services offered by local governments. The following local e-government core indicators have been measured:

EG1 Proportion of persons employed in local government organizations routinely using computers

EG2 Proportion of persons employed in local government organizations routinely using the Internet

EG3 Proportion of local government organizations with a LAN

EG4 Proportion of local government organizations with an intranet

EG5 Proportion of local government organizations with Internet access, by type of access

EG6 Proportion of local government organizations using workflow management and DMS

EG7 Proportion of local government organizations with a web presence

The Survey was conducted in October-November 2013 through direct contact with the persons responsible for IT in the municipalities, while the access to web pages and scanning of the municipalities' offer of e-services was implemented in December 2013.

## 4. ANALYSIS OF RESULTS AND E-READINESS ASSESSMENT

The analysis of the results includes an integral analysis of the entire Pelagonia Region, as well as a comparison between rural and urban municipalities (Table 1).

From the survey it can be concluded that 83.28 percent of municipal employees use a computer while 83.22 percent use the internet while performing daily regular duties. On average, it is almost equal percentage of employees using a computer and/or Internet in rural and urban

International Journal of Managing Public Sector Information and Communication Technologies (IJMPICT)
Vol. 5, No. 2, June 2014communities. This high percentage of users of computer and internet in the performance of daily duties clearly indicates the willingness of the local administration for offering e-services to citizens.

From the answers to the third and fourth question can be concluded that all nine municipalities have installed local area network, while 67 percent have installed intranet (four urban and two rural municipalities). All municipalities have an internet connection where only one rural municipality has narrowband internet access, while 67 percent have fixed broadband access and/or 78 percent wireless broadband internet. It is evident that all urban municipalities have fixed and wireless broadband internet. Regarding the question "Does your units of local self-government use system (software) for managing business processes (workflow management), or system (software) for document management" only three municipalities i.e. 33 percent responded "yes" while the other 6 municipalities gave a negative answer to this question. All three municipalities that responded affirmatively are urban municipalities. In terms of having a website, all municipalities responded positively and pointed to their Internet address.

Table 1. Local e-government readiness in the Pelagonia Region.

| Municipality | EG1 | EG2 | EG3 | EG4 | EG5 | EG5a | EG5b | EG5c | EG6 | EG7 |
|---|---|---|---|---|---|---|---|---|---|---|
| DemirHisar | 57% | 57% | yes | yes | yes | no | yes | yes | no | yes |
| Bitola | 100% | 100% | yes | yes | yes | no | yes | yes | yes | yes |
| Prilep | 100% | 100% | yes | yes | yes | no | yes | yes | yes | yes |
| Resen | 100% | 100% | yes | yes | yes | no | yes | Yes | yes | yes |
| Krusevo | 65% | 65% | yes | no | yes | no | yes | Yes | no | yes |
| Urban Avg. | 84.27% | 84.27% | 100% | 80% | 100% | 0% | 100% | 100% | 60% | 100% |
| Krivogashtani | 83% | 83% | yes | yes | yes | no | yes | no | no | yes |
| Mogila | 76% | 76% | yes | yes | yes | no | no | yes | no | yes |
| Novaci | 79% | 79% | yes | no | yes | no | no | yes | no | yes |
| Dolneni | 89% | 89% | yes | no | yes | yes | no | no | no | yes |
| Rural Avg. | 81.79% | 81.91% | 100% | 50% | 100% | 25% | 25% | 50% | 0% | 100% |
| Total Avg. | 83.28% | 83.22% | 100% | 66.67% | 100% | 11.11% | 66.67% | 77.78% | 33.33% | 100% |

In the second phase, a scan and analysis of the municipality e-services has been done as a key indicator for e-government readiness. The first scan comprised basic services offered by municipalities to their citizens. This scan has set aside the following services as most significant:

1. Access to public information
2. Applications for issuing and tracking building permissions
3. Filing tax forms for property tax
4. Environmental licenses
5. Procedures for issuing municipal utilities approvals
6. Urban planning requirements
7. Requests for local streets and roads
8. Reporting communal problems
9. Ask the Mayor

The values obtained from the survey are presented in Table 2. One can conclude that on a scale of 0 to 4, the lowest score of 0.56 is recorded for the indicator that measures the provision of service for solving problems related to local streets and roads (the average value for the whole Pelagonia region). There is no rural municipality that offers this type of e-service. The highest score of 3.33 is recorded to the "ask the Mayor" indicator. Seven municipalities, out of nine, offer full internet service i.e. they offer interactive service for asking questions, getting answers and solving specific problems which are under the jurisdiction of the Mayor. The values of all measured indicators for the urban and rural communities are presented in Fig. 2.

Table 2. Scores obtained from the survey.

| Muncipality/Indicator | 1 | 2 | 3 | 4 | 5 | 6 | 7 | 8 | 9 | Avg. |
|---|---|---|---|---|---|---|---|---|---|---|
| DemirHisar | 3 | 3 | 1 | 0 | 0 | 2 | 0 | 3 | 4 | 1.78 |
| Bitola | 2 | 4 | 4 | 2 | 2 | 3 | 2 | 1 | 4 | 2.67 |
| Prilep | 2 | 1 | 4 | 2 | 2 | 2 | 2 | 2 | 4 | 2.33 |
| Resen | 2 | 2 | 2 | 2 | 2 | 2 | 1 | 4 | 4 | 2.33 |
| Krusevo | 2 | 2 | 2 | 0 | 2 | 2 | 0 | 3 | 4 | 1.89 |
| Urban (avg) | 2.20 | 2.40 | 2.60 | 1.20 | 1.60 | 2.20 | 1.00 | 2.60 | 4.00 | 2.20 |
| Krivogashtani | 0 | 0 | 1 | 0 | 0 | 0 | 0 | 1 | 1 | 0.33 |
| Mogila | 1 | 0 | 0 | 0 | 0 | 0 | 0 | 0 | 1 | 0.22 |
| Novaci | 1 | 1 | 0 | 0 | 1 | 0 | 0 | 0 | 4 | 0.78 |
| Dolneni | 1 | 2 | 0 | 1 | 0 | 1 | 0 | 0 | 4 | 1.00 |
| RuralAvg. | 0.75 | 0.75 | 0.25 | 0.25 | 0.25 | 0.25 | 0.00 | 0.25 | 2.50 | 0.58 |
| STD | 0.83 | 1.25 | 1.50 | 0.92 | 0.94 | 1.05 | 0.83 | 1.42 | 1.25 | 0.87 |
| Min | 0 | 0 | 0 | 0 | 0 | 0 | 0 | 0 | 1 | 0.22 |
| Max | 3 | 4 | 4 | 2 | 2 | 3 | 2 | 4 | 4 | 2.67 |
| Total (avg) | 1.56 | 1.67 | 1.56 | 0.78 | 1.00 | 1.33 | 0.56 | 1.56 | 3.33 | 1.48 |

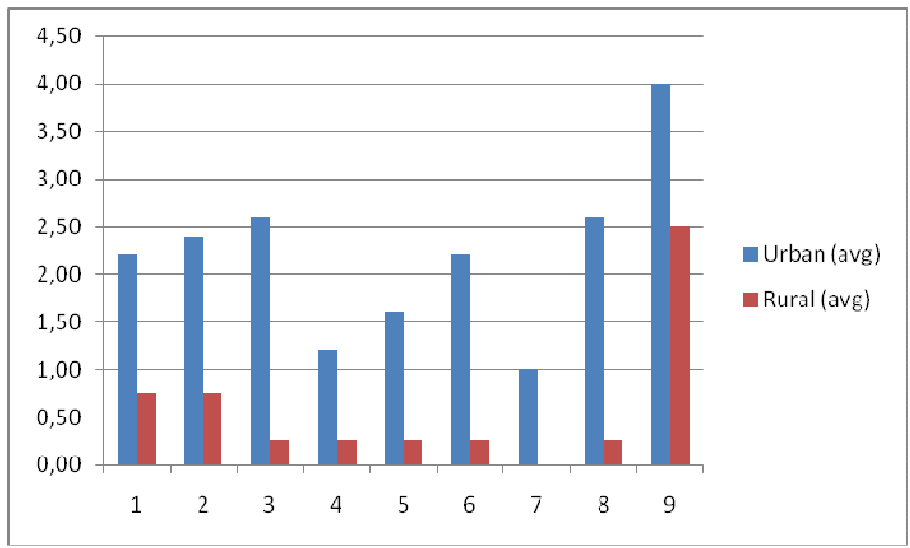

Figure 2. The values of measured indicators for the urban and rural communities

From the calculated values of the standard deviation it can be concluded that the greatest dispersion of data i.e. the widest range of values has the third indicator with SD = 1.5, while the smallest deviation is observed in the first and seventh indicator with the value SD = 0.83.

The average values of measured indicators are much higher in urban than in rural communities. Namely, urban municipalities have a total average index of 2.20 while for the rural municipalities the total average index is 0.58. For the entire Pelagonia region this index is 1.48. The highest index value of 2.67 is recorded in Bitola municipality while the lowest value of 0.22 for municipality of Mogila (Fig. 3).

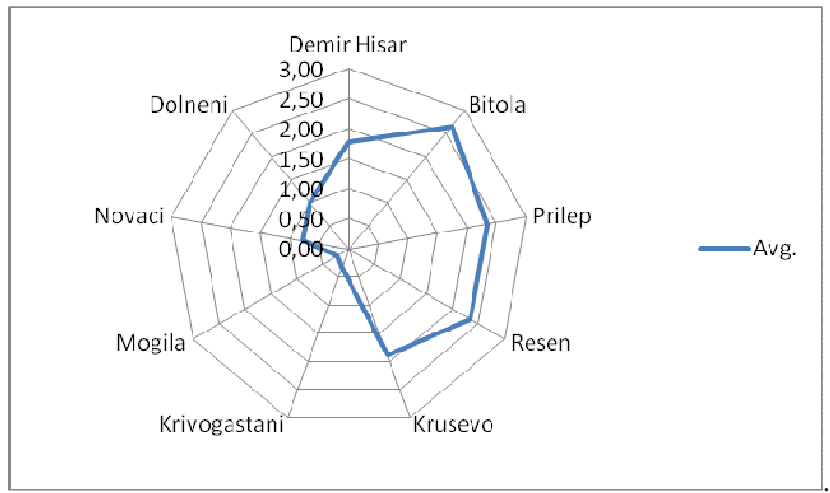

Figure 3. A spider chart representing measured indicators by municipality

## 5. CONCLUSIONS AND RECOMMENDATIONS

Despite the remarkable progress in the supply of local e-services to the citizens, it becomes apparent that they are not sufficiently developed. Namely, the survey revealed that most of the indicators have lower values than 2 on a value scale from 0 to 4. This clearly indicates that the supply of e-services by local authorities is still unsatisfactory and at a relatively low level. Several reasons can be identified for the weaker performance than expected:

- First, although the percentage of internet usage increased more than double since 2005, it is evident that there is an uneven distribution according to the age group of users. The highest percentage of internet usage is recorded among the population belonging to the age group of 15-25 years and the age group of 25 to 54 years. However, the internet usage in the age group over 55 years is relatively low and accounts for only 30 percent. This suggests that many of the people that belong to this age group do not possess computer literacy, while many of those who have registered as Internet users have basic level of knowledge that generates difficulties in using electronic services.

- Second, a large part of the population shows a lack of reliability and confidence for processing their needs and requests electronically. Citizens express more credibility to the traditional methods and practice in obtaining municipal services. In fact they are not sufficiently aware about the benefits gained from the usage of e-services. Even though promotional campaigns were conducted nationwide about the advantages and benefits of e-government, there is an evident need for delivering additional knowledge about e-services that can be obtained at local level.

Hence, local authorities should conduct a series of promotional activities that will bring e-services closer to the citizens and will increase their awareness and trust. Thus the demand for e-services by citizens will rise and will motivate local governments in offering wider range of services.

- Third, local governments have not developed a full set of ICT tools for providing variety of services that can be fully processed electronically. In addition, local governments in rural communities do not possess an adequate technical and administrative capacity to offer e-services with the same quality and volume as the urban municipalities. It is evident that the degree of digitalization in the rural areas is lower than in the urban areas. The rural municipalities do not have sufficient financial resources to cover the initial costs for providing additional technical and infrastructural conditions for installing electronic systems. Despite that the rural municipalities can use funds from various programs of the European Union and other international sources, they do not possess enough knowledge, skills and capacity to develop project applications due to which the degree of utilization of these funds is very low. Likewise, the population from rural areas is characterized with relatively lower average level of education and consequently with insufficient computer skills and knowledge. Given the previous considerations it follows that improving the current situation requires further awareness development among the population pointing that online services can contribute for significant reduction of the costs and time. Therefore, it is necessary to implement a wider delivery of educational contents in terms of methods and techniques for using e-services on one hand, and pointing out the benefits and effects of their usage, on the other.

Regarding long-term perspectives, it is evident that the implementation of the concept of local e-government will be developed with accelerated pace primarily due to the high level of computer literacy among the young people that will naturally increase the demand of e-services provided by local authorities.

**Authors**

**Marjan Angeleski** received his BSc degree from the Faculty of Natural Sciences and Mathematics, Ss. Cyril and Methodius University in Skopje, and MSc and PhD degree in Economics from the Faculty of Economics, St. Clement Ohridski University. His research interests include e-business, innovation and project management. He has published more than 30 papers in journals and conference proceedings. He is currently an assistant professor and Vice-Dean for Education at the Faculty of Economics, St. Clement Ohridski University. 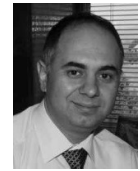

**Pece Mitrevski** received his BSc and MSc degrees in Electrical Engineering and Computer Science, and the PhD degree in Computer Science from the Ss. Cyril and Methodius University in Skopje, Republic of Macedonia. He is currently a full professor and Head of the Department of Computer Science and Engineering at the Faculty of Technical Sciences, St. Clement Ohridski University, Bitola, Republic of Macedonia. His research interests include Computer Architecture, Computer Networks, Performance and Reliability Analysis of Computer Systems, e-Commerce and Behavioral Economics. He has published more than 90 papers in journals and refereed conference proceedings and lectured extensively on these topics. He is a member of the IEEE Computer Society and the ACM. 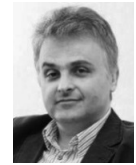

**Slavica Rocheska** received her MSc degree in International Business Law from the Ss. Cyril and Methodius University in Skopje and PhD degree in Economics from the Faculty of Economics, St. Clement Ohridski University in Bitola, Republic of Macedonia. Currently she works as a full professor at the Faculty of Economics in Prilep. Her research interests include Innovation, International Economics and International Business. She has published three books and more than 80 papers in journals and refereed conference proceedings. 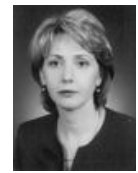

**Ane Lashkoska** graduated from the Faculty of Natural Sciences and Mathematics, Ss. Cyril and Methodius University in Skopje, Republic of Macedonia. She is currently a MSc student at the Faculty of Economics in Prilep, St. Clement Ohridski University. Her field of interest is e-Government. From 2011 she is a member of the Macedonian Parliament and deputy-member of the Committee on Equal Opportunities for Women and Men. 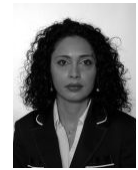